# Gigahertz acousto-optic modulation and frequency shifting on etchless lithium niobate integrated platform

Zejie Yu and Xiankai Sun[*]

*Department of Electronic Engineering, The Chinese University of Hong Kong, Shatin, New Territories, Hong Kong SAR, China*

[*]*Corresponding author: xksun@cuhk.edu.hk*

**Abstract**

Acousto-optic interactions involving propagating phonons can break the time-reversal and frequency-modulation symmetry of light. However, conventional acousto-optic modulators based on bulk materials have frequency bandwidth limited to hundreds of megahertz due to their large structural sizes. Here, we experimentally demonstrate gigahertz single-sideband acousto-optic modulation on an etchless lithium niobate integrated platform by using photonic bound states in the continuum. The upper- or lower-sideband modulation of light can be obtained conveniently by choosing specific combinations of input and output channels. Under this scheme, we have realized a 3-GHz frequency shifter, which operates in the C-band with a 3-dB bandwidth of ~35 nm. The extinction ratios of the upper(lower)-sideband modulated light to the lower(upper)-sideband modulated and unmodulated light are >44 (47) and 25 (23) dB in the 3-dB operating bandwidth. The frequency-shifted light can be further processed with amplitude and frequency modulation. Therefore, the demonstrated gigahertz single-sideband acousto-optic modulation can enable many photonic applications such as optical signal processing, sensing, and ion trapping.

**Introduction**

Acousto-optics, such as Brillouin scattering[1-9], involves the study of phonon–photon interactions based on changes in the refractive index of a medium due to the presence of acoustic waves in that medium. Because the interactions involve phonons with a momentum, such processes require satisfaction of the phase-matching condition and can break the time-reversal and frequency-modulation symmetry of light. As a result, acousto-optic (AO) interactions can be utilized to achieve nonreciprocal and single-sideband modulation of light, which has applications in many areas such as nonreciprocal light transmission[10-15], modulation[16], frequency shifting[17], and signal processing[18-20]. Presently, single-sideband AO interactions are facing a bottleneck in enhancing the modulation frequency because of the large structural size of devices based on bulk materials. Actually, the AO modulators commercially available in the market usually an operating frequency limited within hundreds of megahertz. AO modulators operating at gigahertz frequencies are highly desired in various applications. Therefore, there is an urgent demand for single-sideband AO interactions in the gigahertz frequency range.

Surface acoustic waves (SAWs) that propagate on surfaces of a thin-film piezoelectric material can be confined within a thickness less than the acoustic wavelength, producing phonons with a very high density in the region near the surface. The small acoustic modal area, which is comparable to the optical modal area, can result in a large overlap between the two modes in photonic waveguides. Therefore, SAWs can be used to achieve strong AO interactions in nanophotonic devices[16,21-25]. In addition, SAWs can be excited electromechanically, with the acoustic frequency reaching tens of gigahertz in piezoelectric materials. Lithium niobate ($LiNbO_3$) has large piezoelectric coefficients and is optically transparent over a wide wavelength range. It can be used to generate SAWs efficiently and support low-loss optical waveguides. Therefore, thin-film $LiNbO_3$ is an ideal platform for realizing single-sideband AO interactions in the gigahertz frequency range[24,26,27].

Here, we experimentally demonstrated single-sideband AO modulations of light on a thin-film $LiNbO_3$-on-insulator platform. Photonic bound states in the continuum (BICs) were harnessed for low-loss optical waveguiding[28-32]. The BIC-based optical waveguides were constructed simply by patterning a low-refractive-index material on the high-refractive-index $LiNbO_3$ substrate without facing the challenge of high-quality etching of $LiNbO_3$[31,32]. The SAWs were electrically excited via a monolithically integrated SAW interdigital transducer (IDT), and propagated smoothly in the unetched LiNbO$_3$ thin film. The demonstrated AO modulations of light are single-sidebanded in the entire frequency bandwidth of the acoustic waves. The upper- or lower-sideband modulation of light can be obtained conveniently by choosing specific combinations of input and output channels. Under this scheme, we realized a 3-GHz frequency shifter, which operates in the C-band with a 3-dB bandwidth of ~35 nm and has convenient control of the frequency shifting direction. The extinction ratios of the upper(lower)-sideband modulated light to the lower(upper)-sideband modulated and unmodulated light are >44 (47) and 25 (23) dB in the 3-dB operating bandwidth. The highest efficiency for the frequency up(down)-shifting is ~2.8(2.4)%/W. In addition, we further experimentally demonstrated both amplitude modulation and frequency modulation with the fabricated frequency shifter.

**Results**

Figure 1a is a schematic illustration of the device for investigating single-sideband AO modulations, where an IDT is monolithically integrated with photonic waveguides that support the BIC modes on a thin-film LiNbO$_3$ platform. The yellow part denotes the IDT patterned in gold (Au), the purple part denotes the low-refractive-index waveguides patterned in polymer (electron-beam resist ZEP520A), the pink part denotes the high-refractive-index LiNbO$_3$ thin film, and the gray part denotes silicon oxide (SiO$_2$). The device has four different channels labeled as $C_1$–$C_4$, and $\theta$ denotes the angle between channels $C_1$ ($C_3$) and $C_2$ ($C_4$). The close-up shown on the right is a top view of part of the device, where $\Lambda$, $d$, and $D$ denote the period, finger width, and aperture of the IDT, respectively. For light input from channel $C_1$, its |**E**| field profile in the (TM-polarized) BIC mode in the waveguide and the free-propagation region at the positions marked in the close-up in Fig. 1a is plotted in Figs. 1b and 1c, respectively. It is clear that light can be guided and laterally confined in the LiNbO$_3$ thin film to have strong interactions with SAWs. Figures 1d and 1e illustrate the sideband modulation processes for light input from channel $C_1$ and $C_2$, respectively.

The SAWs excited by the IDT propagate in the LiNbO$_3$ thin film along the direction indicated by the thick black arrows. Light input from channel $C_1$ ($C_2$) interacts with the counterpropagating (copropagating) traveling SAWs. Because of the Doppler effect of light (see Sec. 1 of the Supplementary Information), the light scattered to the first order should have a frequency up(down)-shift. The scattered light beam is deflected from the original light beam by an angle $\theta$, and thus is coupled into channel $C_3$ ($C_4$) for output. The unscattered portion of the original light beam maintains its propagation direction, and thus is coupled into channel $C_4$ ($C_3$). Therefore, it is convenient to obtain the upper- or lower-sideband AO modulation of light by choosing specific combinations of input and output channels.

We fabricated the devices on a 400-nm $z$-cut LiNbO$_3$-on-insulator wafer with silicon as the substrate handle. Figure 2a shows an optical microscope image of the fabricated device. The angle between channels $C_1$ ($C_3$) and $C_2$ ($C_4$) is 30°. The close-up on the right is a scanning electron microscope (SEM) image showing the details of the IDT, whose period, finger width, and aperture are 1.81 μm, 452 nm, and 75 μm, respectively. The period of the IDT was determined by the phase-matching condition in the free-propagation region, where the effective refractive index of the TM mode is ~1.65 at the wavelength of 1.55 μm. In the free-propagation region, the traveling SAWs excited by the IDT cause Brillouin scattering to the light, resulting in upper- or lower-sideband modulation. Outside the free-propagation region, the LiNbO$_3$ thin film is covered by the unpatterned polymer. It should be noted that the SAWs reaching the boundary of the free-propagation region would experience negligible reflection, because the polymer atop has very different acoustic properties from those of LiNbO$_3$. The transmitted SAWs would be totally absorbed by the unpatterned polymer outside the free-propagation region. Therefore, the SAWs traveling in the free-propagation region can have ultrahigh purity, because of elimination of the unwanted reflected SAWs, which contributes to the high quality of single-sideband AO modulation.

We measured the AO interaction of our devices by using the experimental setup shown in Fig. 2b. Light from a tunable semiconductor laser was split into two branches by a 90:10 fiber coupler. In the first branch, the light with 90% of power was sent through a fiber polarization controller to adjust its polarization state before being sent into the device under test. The light was coupled into and out of the device via a pair of grating couplers, because the grating couplers are polarization sensitive and thus facilitate high-efficiency excitation of the fundamental TM-polarized BIC mode in the on-chip waveguides[33]. Meanwhile, a sinusoidal microwave signal at the frequency of $\Omega$ from a signal generator was delivered to the IDT of the device via a microwave probe to excite the SAWs in the LiNbO$_3$ thin film. In the second branch, the light with 10% of power was sent into a commercial fiber-coupled AO modulator with a frequency shift of $\Omega_{ao}$ = 200 MHz. The acousto-optically modulated light coupled out of our fabricated device in the first branch was combined with the frequency-shifted light in the second branch by a 50:50 fiber coupler, after which the light was collected by a high-speed photodetector and then sent into an electrical signal analyzer (ESA), for analysis in the electrical domain. With this heterodyne detection scheme, we could distinguish the upper and lower sidebands from AO modulation, and measure their individual strengths with sufficiently high resolution and sensitivity. The signals measured from the ESA at the frequency of $\Omega - \Omega_{ao}$ ($\Omega + \Omega_{ao}$) correspond to the upper (lower) sideband of the AO modulation in our fabricated device. Figures 2c and 2d plot the electrical power spectra measured from the ESA with the light wavelength at 1.54 μm and the light input–output channels being $C_1$–$C_3$ and $C_2$–$C_4$, respectively. It is clear that the fabricated device can achieve predominantly upper- or lower-sideband modulation, with a high extinction ratio between the two sidebands in the entire measured frequency range of 1–4 GHz. The results in Figs. 2c and 2d also show that the upper- or lower-sideband modulation of light can be obtained conveniently by choosing specific combinations of input and output channels. The single-sideband modulation results for other wavelengths are shown in Sec. 2 of the Supplementary Information.

Because singe-sideband modulation of light can be produced in the free-propagation region, and only the modulated portion of light input from channel $C_1$ ($C_2$) is deflected into channel $C_3$ ($C_4$) while the unmodulated portion propagates along its original direction to channel $C_4$ ($C_3$), our fabricated device can work as an integrated frequency shifter capable of shifting the frequency of the input light by several gigahertz. In order to achieve both high efficiency and large extinction ratio in frequency shifting, we set the microwave driving frequency at 3 GHz for the fabricated device to achieve a 3-GHz frequency shift. Figures 3a and 3b plot the measured electrical power spectra showing the upper-sideband modulated, unmodulated, and lower-sideband modulated light at the wavelength of 1.54 μm with the light input–output channels being $C_1$–$C_3$ and $C_2$–$C_4$, respectively. The signal strengths were obtained from the ESA at the frequencies of 2.8 GHz, 200 MHz, and 3.2 GHz, respectively. The frequency up(down)-shifting can achieve extinction ratios of the upper(lower)-sideband modulated light to the lower(upper)-sideband modulated and unmodulated light as large as 49 (51) and 27 (26) dB, respectively. Figures 3c and 3d plot the measured signal strength for the upper-sideband modulated (dots), unmodulated (crosses), and lower-sideband modulated (triangles) light in the wavelength range of 1.51–1.57 μm with the light input–output channels being $C_1$–$C_3$ and $C_2$–$C_4$, respectively. The 3-dB bandwidth for both frequency up- and down-shifting is as large as ~35 nm. The extinction ratios of the upper(lower)-sideband modulated light to the lower(upper)-sideband modulated and unmodulated light are >44 (47) and 25 (23) dB in the 3-dB operating bandwidth. In our experiment, we found that the measured unmodulated light was mainly from undesired scattering in the device and direct coupling between the input and output fibers, because the unmodulated light maintained the same strength no matter whether the microwave driving signal was applied on the IDT. The measured undesired sideband modulations were mainly caused by the imperfect SAWs especially at their edges. In addition, the measured frequency shifting efficiency is ~2.8%/W and ~2.4%/W for the upper and lower sideband, respectively, which can significantly be enhanced by improving the design of the IDT electrodes (see Sec. 2 of the Supplementary Information).

Last, we performed amplitude modulation (AM) and frequency modulation (FM) with the fabricated frequency shifter. As the frequency shifting efficiency is highly dependent on the power and frequency of the microwave driving signal, the power of the frequency-shifted light can be modulated through AM and FM. Figure 4a shows the experimental setup for the AM and FM measurements. Light from a tunable semiconductor laser was sent through a fiber polarization controller to adjust its polarization state before being coupled into channel $C_2$ of the fabricated device. Meanwhile, a microwave signal from a signal generator was delivered to the IDT of the fabricated device to modulate the light. The microwave signal with a carrier frequency of 3 GHz was either amplitude modulated or frequency modulated, as illustrated by the blue waveforms in Fig. 4a. The modulated light was coupled out of channel $C_4$ of the device and then collected by a photodetector. The electrical signals from the photodetector were sent to an oscilloscope for monitoring. Figure 4b plots the measured optical power of light which had a 3-GHz frequency shift with 50% (red) and 100% (black) AM depth. Figure 4c plots the measured optical power of light which had a 3-GHz frequency shift with 10-MHz (red) and 20-MHz (black) FM. The modulation speed of both the AM and FM was 1 kHz. These results have clearly shown that the frequency-shifted light can be processed further through both AM and FM.

**Discussion**

In conclusion, we have experimentally demonstrated gigahertz single-sideband acousto-optic modulation, with a high extinction ratio between the modulation strengths of the desired and undesired sidebands, on an etchless lithium niobate integrated platform by using photonic bound states in the continuum. The upper- or lower-sideband modulation of light can be obtained conveniently by choosing specific combinations of input and output channels. Harnessing these features, we further demonstrated a 3-GHz acousto-optic frequency shifter, which operates in the C-band with a 3-dB bandwidth of ~35 nm for both frequency up- and down-shifting. The frequency shifting direction is associated with specific combinations of input and output channels and thus can be controlled conveniently. The extinction ratios of the upper(lower)-sideband modulated light to the lower(upper)-sideband modulated and unmodulated light are >44 (47) and 25 (23) dB in the 3-dB operating bandwidth. The highest efficiency for the frequency up(down)-shifting is ~2.8(2.4)%/W. The frequency-shifted light can be further processed by amplitude modulation and frequency modulation. Therefore, the gigahertz single-sideband modulation demonstrated in this work can be harnessed for developing a wide range of Brillouin-scattering-based photonic applications, including ion trapping, microwave signal processing, Brillouin lasers and amplifiers, and quantum information.

**Materials and methods**

The devices were fabricated on a *z*-cut $LiNbO_3$-on-insulator wafer purchased from NANOLN, where the nominal thickness of the $LiNbO_3$ layer is 400 nm. We first fabricated the IDTs (titanium/gold with thicknesses of 5/100 nm) with a lift-off process involving electron-beam lithography and metal deposition, where titanium was used as an adhension layer. Then, we performed a second step of electron-beam lithography to pattern the photonic waveguides and grating couplers in an electron-beam resist (ZEP520A), which serves as the polymer layer in Fig. 1a. The thickness of the electron-beam resist ZEP520A was controlled to be ~500 nm by using a spinning speed of 2400 r/min during spin coating.

**Acknowledgments**

This work was supported by the General Research Fund (14208717, 14206318, 14209519) sponsored by the Research Grants Council of Hong Kong, and by the NSFC/RGC Joint Research Scheme (N_CUHK415/15) sponsored by the Research Grants Council of Hong Kong and the National Natural Science Foundation of China.

**Conflict of interests**

The authors declare no competing interests.

**Author contributions**

Z.Y. performed the theoretical modeling, numerical simulation, device design, fabrication, and measurement under the supervision of X.S.; Z.Y. and X.S. wrote the paper.

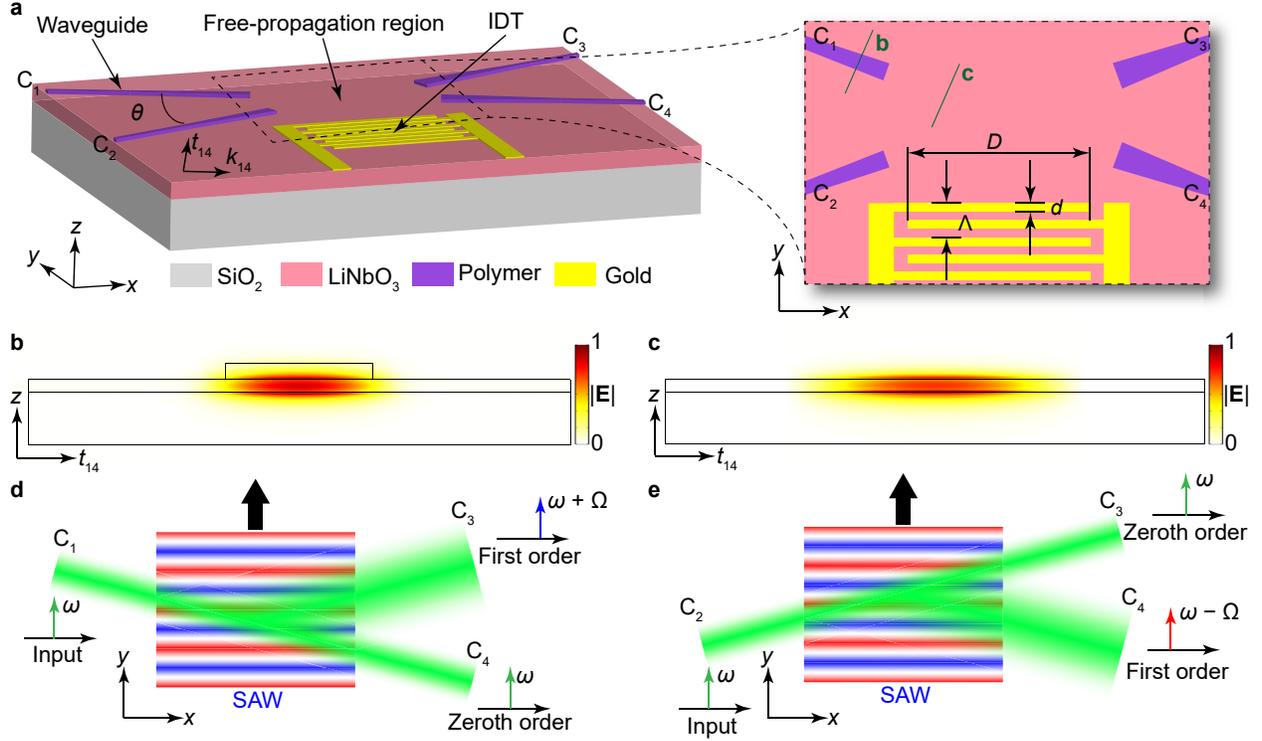

**Fig. 1. Scheme of single-sideband modulation on etchless lithium niobate with photonic bound state in the continuum. a** Schematic illustration of the device structure. $k_{14}$ and $t_{14}$ denote the longitudinal and transverse directions of the waveguides in channels $C_1$ and $C_4$. The close-up on the right is a top view of part of the device, where $\Lambda$, $d$, and $D$ denote the period, finger width, and aperture of the IDT, respectively. **b**, **c** Simulated modal profiles of electric field $|\mathbf{E}|$ in the waveguide (**b**) and the free-propagation region (**c**) as marked in the close-up of **a**. **d**, **e** Upper- (**d**) and lower- (**e**) sideband modulation processes for light input from channel $C_1$ (**d**) and $C_2$ (**e**), respectively.

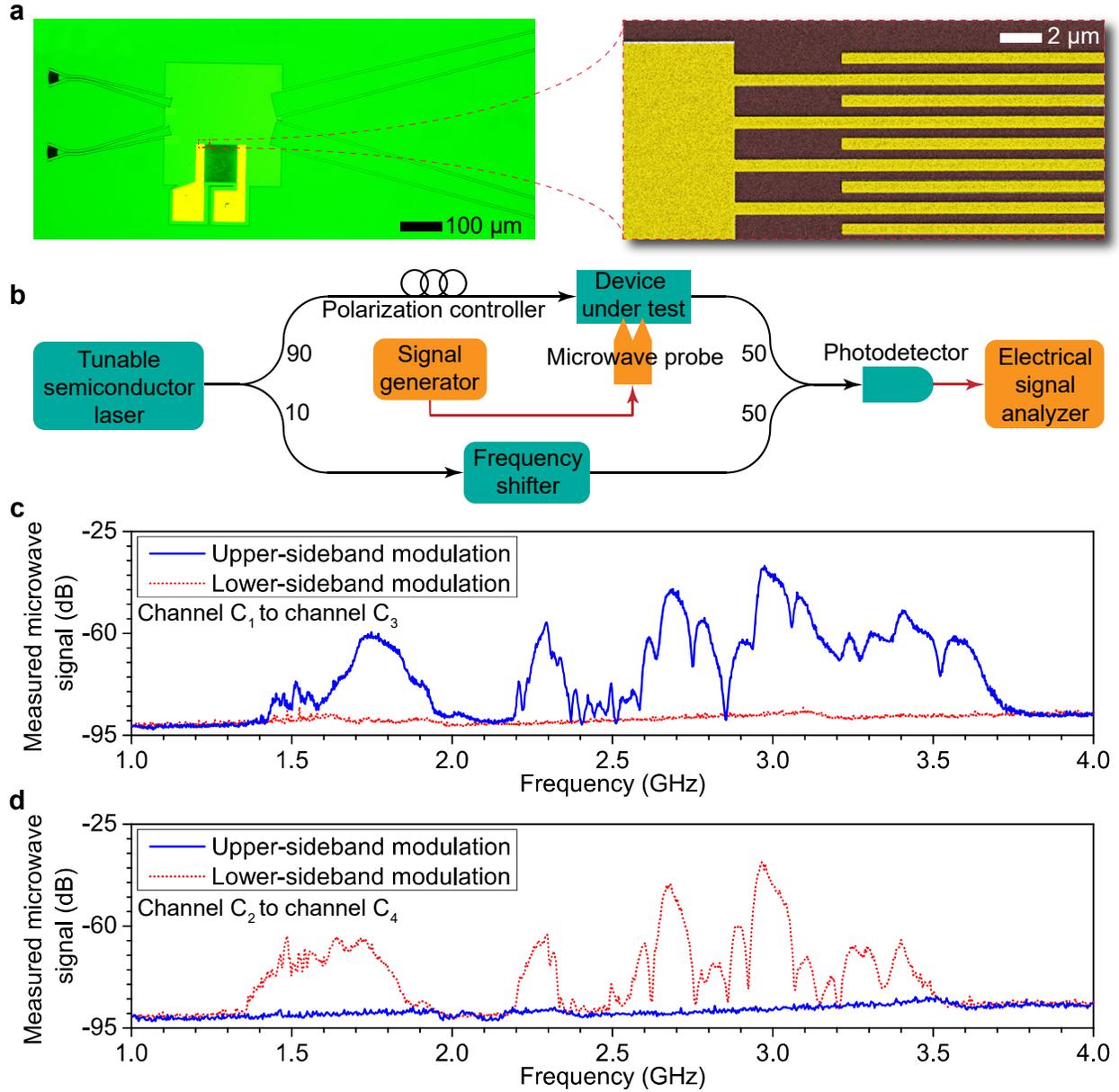

**Fig. 2. Experimental demonstration of single-sideband modulations. a** Optical microscope image of the fabricated device. The close-up on the right is an SEM image showing the details of the IDT. **b** Experimental setup for measuring the modulation strength of both the upper and lower sidebands simultaneously. **c**, **d** Upper- (solid) and lower- (dotted) sideband modulation strength at the wavelength of 1.54 μm measured with the light input–output channels being $C_1$–$C_3$ (**c**) and $C_2$–$C_4$ (**d**).

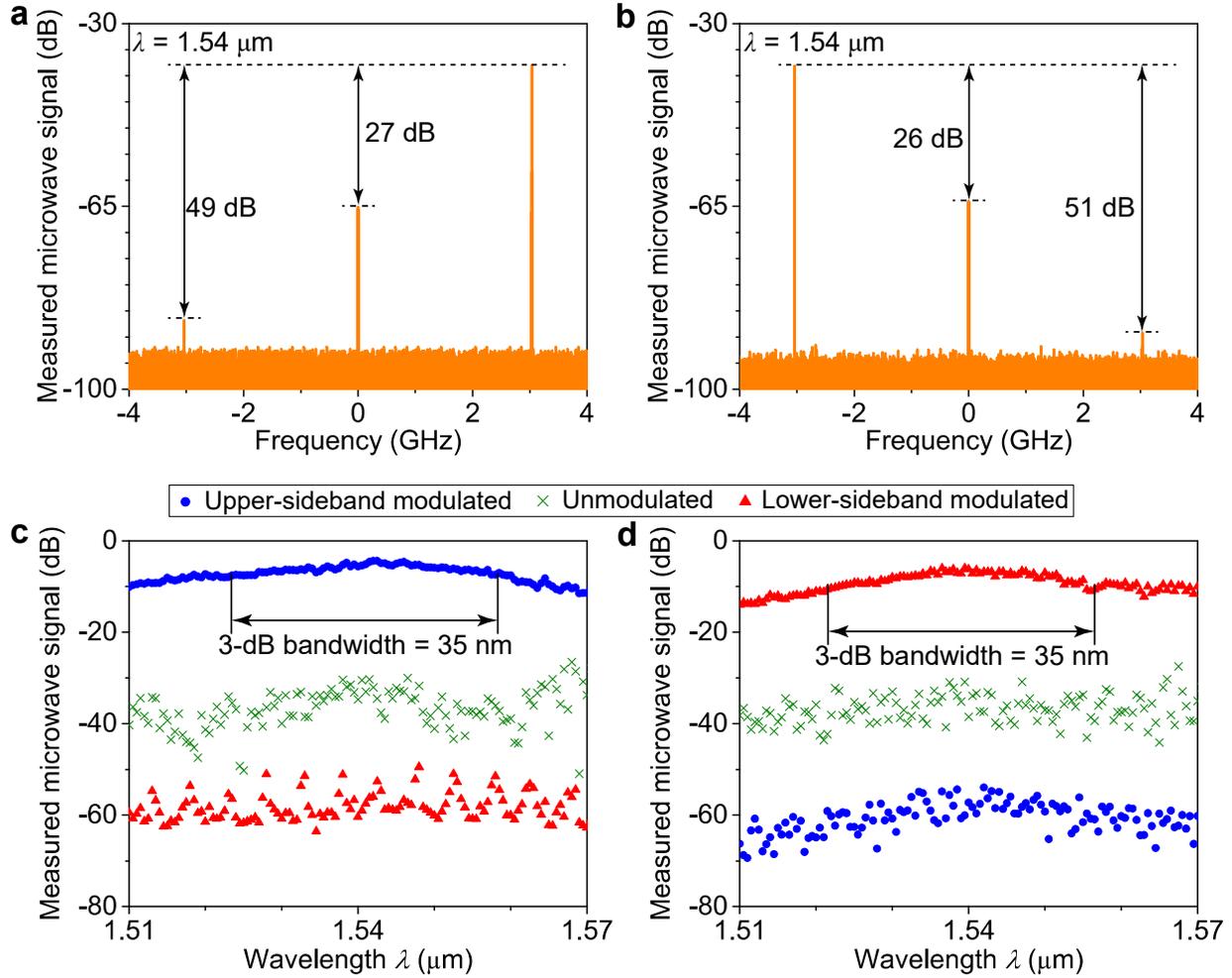

**Fig. 3. Experimental demonstration of frequency shifting. a**, **b** Electrical power spectra showing the upper-sideband modulated, unmodulated, and lower-sideband modulated light measured at the wavelength of 1.54 μm with the light input–output channels being $C_1$–$C_3$ (**a**) and $C_2$–$C_4$ (**b**). **c**, **d** Signal strength for the upper-sideband modulated (dots), unmodulated (crosses), and lower-sideband modulated (triangles) light measured in the wavelength range of 1.51–1.57 μm with the light input–output channels being $C_1$–$C_3$ (**c**) and $C_2$–$C_4$ (**d**). The 3-dB bandwidth for both frequency up- and down-shifting is as large as 35 nm.

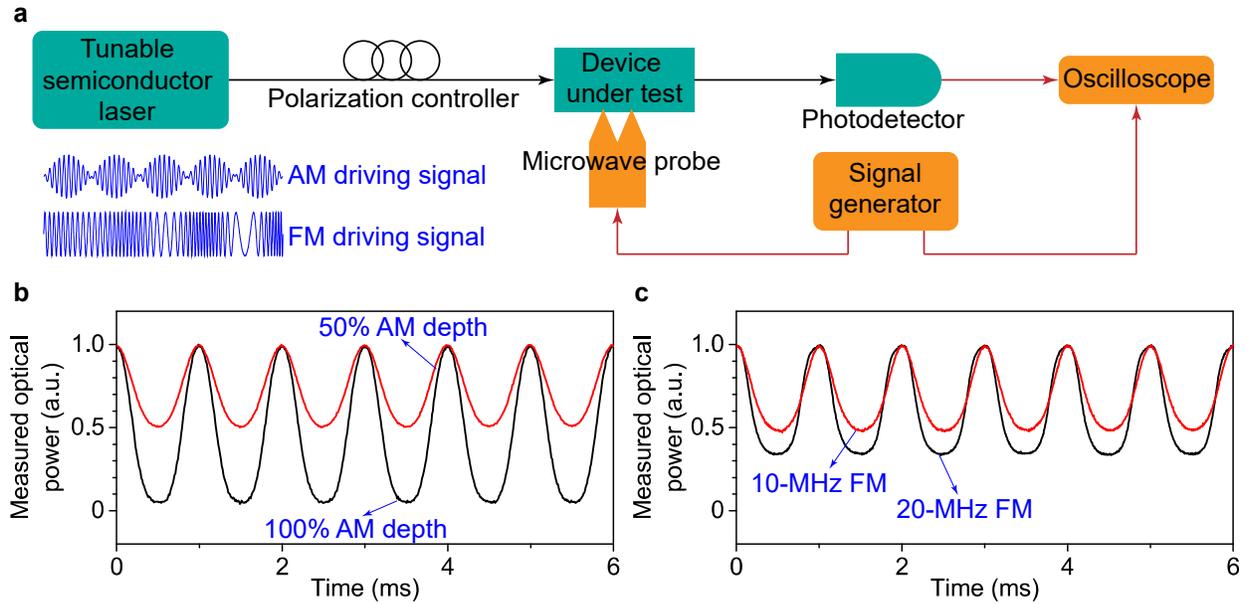

**Fig. 4. Experimental demonstration of AM and FM with the fabricated frequency shifter. a** Experimental setup for measuring AM and FM with the fabricated frequency shifter. The AM or FM driving signals (represented by the blue waveforms) are delivered from a signal generator to the device under test via a microwave probe. **b** Measured optical power of light with a 3-GHz frequency shift under 50% (red) and 100% (black) AM depth. **c** Measured optical power of light with a 3-GHz frequency shift under 10-MHz (red) and 20-MHz (black) FM. The modulation speed of both the AM and FM is 1 kHz.